%% file: main.tex
\pdfoutput=1

\documentclass[12pt,a4paper]{article}

\usepackage{ifthen} 
\usepackage{multirow}
\usepackage{enumitem} 
\usepackage{float} 
\usepackage{amsmath}
\newboolean{pdflatex}
\setboolean{pdflatex}{true} 

\newboolean{articletitles}
\setboolean{articletitles}{true} 

\newboolean{uprightparticles}
\setboolean{uprightparticles}{false} 
\newcommand{\Yn}[1]{\ensuremath{\PUpsilon{(#1S)}\xspace}}
\def\pvt{N_{\mathrm{trk}}^{\mathrm{PV}}}
\def\bkt{N_{\mathrm{bwd}}^{\mathrm{PV}}}
\def\fwt{N_{\mathrm{fwd}}^{\mathrm{PV}}}

\def\paperauthors{LHCb collaboration} 
\def\paperasciititle{Measurement of the multiplicity dependence of Upsilon production ratios in pp collisions at centre of mass energy 13 TeV$} 
\def\papertitle{Measurement of the multiplicity dependence of $\PUpsilon$ production ratios in $pp$ collisions at $\sqrt{s}=13\tev$} 
\def\paperkeywords{{High Energy Physics}, {LHCb}} 
\def\papercopyright{\the\year\ CERN for the benefit of the LHCb collaboration} 
\def\paperlicence{CC BY 4.0 licence}
\def\paperlicenceurl{https://creativecommons.org/licenses/by/4.0/}

\input{preamble}
\usepackage{longtable} 

\begin{document}

\renewcommand{\thefootnote}{\fnsymbol{footnote}}
\setcounter{footnote}{1}

\input{title-LHCb-PAPER}


\renewcommand{\thefootnote}{\arabic{footnote}}
\setcounter{footnote}{0}

\cleardoublepage


\pagestyle{plain} 
\setcounter{page}{1}
\pagenumbering{arabic}

\input{body}

\input{acknowledgements}

\newpage


\addcontentsline{toc}{section}{References}
\bibliographystyle{LHCb}
\bibliography{main,standard,LHCb-PAPER,LHCb-CONF,LHCb-DP,LHCb-TDR}

\newpage
\input{Authorship_LHCb-PAPER-2024-038}

\end{document}

%% file: preamble.tex
\usepackage[top=1in, bottom=1.25in, left=1in, right=1in]{geometry}

\usepackage{microtype}
\usepackage{lineno}  
\usepackage{xspace} 
\usepackage{caption} 

\usepackage{graphicx}  
\usepackage{color}
\usepackage{colortbl}
\graphicspath{{./figs/}} 

\usepackage{amsmath} 
\usepackage{amssymb}
\usepackage{amsfonts}
\usepackage{upgreek} 

\newcommand*\patchAmsMathEnvironmentForLineno[1]{%
\expandafter\let\csname old#1\expandafter\endcsname\csname #1\endcsname
\expandafter\let\csname oldend#1\expandafter\endcsname\csname
end#1\endcsname
 \renewenvironment{#1}%
   {\linenomath\csname old#1\endcsname}%
   {\csname oldend#1\endcsname\endlinenomath}%
}
\newcommand*\patchBothAmsMathEnvironmentsForLineno[1]{%
  \patchAmsMathEnvironmentForLineno{#1}%
  \patchAmsMathEnvironmentForLineno{#1*}%
}
\AtBeginDocument{%
\patchBothAmsMathEnvironmentsForLineno{equation}%
\patchBothAmsMathEnvironmentsForLineno{align}%
\patchBothAmsMathEnvironmentsForLineno{flalign}%
\patchBothAmsMathEnvironmentsForLineno{alignat}%
\patchBothAmsMathEnvironmentsForLineno{gather}%
\patchBothAmsMathEnvironmentsForLineno{multline}%
\patchBothAmsMathEnvironmentsForLineno{eqnarray}%
}

\usepackage{hyperxmp}

\usepackage[pdftex,
            pdfauthor={\paperauthors},
            pdftitle={\paperasciititle},
            pdfkeywords={\paperkeywords},
            pdfcopyright={Copyright (C) \papercopyright},
            pdflicenseurl={\paperlicenceurl}]{hyperref}

\usepackage[colorinlistoftodos,textsize=scriptsize]{todonotes}

\usepackage[bottom,flushmargin,hang,multiple]{footmisc}

\usepackage[all]{hypcap} 

\input{lhcb-symbols-def} 

\usepackage{cite} 
\usepackage{mciteplus}

%% file: lhcb-symbols-def.tex
\usepackage{xspace} 
\usepackage{upgreek}

\def\MagUp {\mbox{\em Mag\kern -0.05em Up}\xspace}

\ifthenelse{\boolean{uprightparticles}}%
{

 \def\Pmu         {\ensuremath{\upmu}\xspace}

 \def\Ppsi        {\ensuremath{\uppsi}\xspace}

 \def\PDelta      {\ensuremath{\Delta}\xspace}                 
 \def\PXi         {\ensuremath{\Xi}\xspace}                 
 \def\PLambda     {\ensuremath{\Lambda}\xspace}                 
 \def\PSigma      {\ensuremath{\Sigma}\xspace}                 
 \def\POmega      {\ensuremath{\Omega}\xspace}                 
 \def\PUpsilon    {\ensuremath{\Upsilon}\xspace}
 \let\oldPi\Pi
 \def\PPi         {\ensuremath{\oldPi}\xspace}

 \def\PB      {\ensuremath{\mathrm{B}}\xspace}                 
                  
 \def\PD      {\ensuremath{\mathrm{D}}\xspace}

 \def\PJ      {\ensuremath{\mathrm{J}}\xspace}                 
 \def\PK      {\ensuremath{\mathrm{K}}\xspace}

 \def\Pi      {\ensuremath{\mathrm{i}}\xspace}

 \def\Ps      {\ensuremath{\mathrm{s}}\xspace}

 \def\thebaroffset{0.0em}
}
{

 \def\Pmu         {\ensuremath{\mu}\xspace}

 \def\Ppsi        {\ensuremath{\psi}\xspace}                 
                  
 \mathchardef\PDelta="7101
 \mathchardef\PXi="7104
 \mathchardef\PLambda="7103
 \mathchardef\PSigma="7106
 \mathchardef\POmega="710A
 \mathchardef\PUpsilon="7107
 \mathchardef\PPi="7105
                  
 \def\PB      {\ensuremath{B}\xspace}                 
                  
 \def\PD      {\ensuremath{D}\xspace}

 \def\PJ      {\ensuremath{J}\xspace}                 
 \def\PK      {\ensuremath{K}\xspace}

 \def\Pi      {\ensuremath{i}\xspace}

 \def\Ps      {\ensuremath{s}\xspace}

 \def\thebaroffset{0.18em}
}
\newcommand{\offsetoverline}[2][\thebaroffset]{\kern #1\overline{\kern -#1 #2}}%

\makeatletter
\ifcase \@ptsize \relax
  \newcommand{\miniscule}{\@setfontsize\miniscule{4}{5}}
\or
  \newcommand{\miniscule}{\@setfontsize\miniscule{5}{6}}
\or
  \newcommand{\miniscule}{\@setfontsize\miniscule{5}{6}}
\fi
\makeatother

\DeclareRobustCommand{\optbar}[1]{\shortstack{{\miniscule (\rule[.5ex]{1.25em}{.18mm})}
  \\ [-.7ex] $#1$}}

\def\mup        {{\ensuremath{\Pmu^+}}\xspace}
\def\mun        {{\ensuremath{\Pmu^-}}\xspace} 

\def\mumu       {{\ensuremath{\Pmu^+\Pmu^-}}\xspace}

\def\squark    {{\ensuremath{\Ps}}\xspace}

\def\KorKbar {\kern \thebaroffset\optbar{\kern -\thebaroffset \PK}{}\xspace}

\def\D       {{\ensuremath{\PD}}\xspace}

\def\DorDbar {\kern \thebaroffset\optbar{\kern -\thebaroffset \PD}\xspace}

\def\Dp      {{\ensuremath{\D^+}}\xspace}
\def\Dm      {{\ensuremath{\D^-}}\xspace}

\def\DpDm    {\ensuremath{\Dp {\kern -0.16em \Dm}}\xspace}

\def\B       {{\ensuremath{\PB}}\xspace}

\def\BorBbar {\kern \thebaroffset\optbar{\kern -\thebaroffset \PB}\xspace}

\def\Bd      {{\ensuremath{\B^0}}\xspace}

\def\BdorBdbar {\kern \thebaroffset\optbar{\kern -\thebaroffset \Bd}\xspace}

\def\Bs      {{\ensuremath{\B^0_\squark}}\xspace}

\def\BsorBsbar {\kern \thebaroffset\optbar{\kern -\thebaroffset \Bs}\xspace}

\def\jpsi     {{\ensuremath{{\PJ\mskip -3mu/\mskip -2mu\Ppsi}}}\xspace}
\def\psitwos  {{\ensuremath{\Ppsi{(2S)}}}\xspace}

\def\Y#1S{\ensuremath{\PUpsilon{(#1S)}}\xspace}

\def\LorLbar     {\kern \thebaroffset\optbar{\kern -\thebaroffset \PLambda}\xspace}

\def\BF         {{\ensuremath{\mathcal{B}}}\xspace}
\def\BR         {\BF}

\def\to                 {\ensuremath{\rightarrow}\xspace}

\newcommand{\etot}{{\ensuremath{\varepsilon_{\mathrm{ tot}}}}\xspace}

\def\AT#1     {\ensuremath{A_{\mathrm{T}}^{#1}}\xspace}           

\def\C#1      {\ensuremath{\mathcal{C}_{#1}}\xspace}                       
\def\Cp#1     {\ensuremath{\mathcal{C}_{#1}^{'}}\xspace}                    
\def\Ceff#1   {\ensuremath{\mathcal{C}_{#1}^{\mathrm{(eff)}}}\xspace}        
\def\Cpeff#1  {\ensuremath{\mathcal{C}_{#1}^{'\mathrm{(eff)}}}\xspace}       
\def\Ope#1    {\ensuremath{\mathcal{O}_{#1}}\xspace}                       
\def\Opep#1   {\ensuremath{\mathcal{O}_{#1}^{'}}\xspace}                    

\newcommand{\nospaceunit}[1]{\ensuremath{\text{#1}}}       
\newcommand{\aunit}[1]{\ensuremath{\text{\,#1}}}       

\newcommand{\tev}{\aunit{Te\kern -0.1em V}\xspace}
\newcommand{\gev}{\aunit{Ge\kern -0.1em V}\xspace}
\newcommand{\mev}{\aunit{Me\kern -0.1em V}\xspace}
\newcommand{\kev}{\aunit{ke\kern -0.1em V}\xspace}
\newcommand{\ev}{\aunit{e\kern -0.1em V}\xspace}
 
\newcommand{\mevc}{\ensuremath{\aunit{Me\kern -0.1em V\!/}c}\xspace}
\newcommand{\gevc}{\ensuremath{\aunit{Ge\kern -0.1em V\!/}c}\xspace}
\newcommand{\mevcc}{\ensuremath{\aunit{Me\kern -0.1em V\!/}c^2}\xspace}
\newcommand{\gevcc}{\ensuremath{\aunit{Ge\kern -0.1em V\!/}c^2}\xspace}

\def\mum  {\ensuremath{\,\upmu\nospaceunit{m}}\xspace}

\def\fb   {\ensuremath{\aunit{fb}}\xspace}
\def\invfb   {\ensuremath{\fb^{-1}}\xspace}

\newcommand{\chisqndf}{\ensuremath{\chi^2/\mathrm{ndf}}\xspace}

\def\deriv {\ensuremath{\mathrm{d}}}

\def\gsim{{~\raise.15em\hbox{$>$}\kern-.85em
          \lower.35em\hbox{$\sim$}~}\xspace}
\def\lsim{{~\raise.15em\hbox{$<$}\kern-.85em
          \lower.35em\hbox{$\sim$}~}\xspace}

\def\sPlot{\mbox{\em sPlot}\xspace}

\def\pt         {\ensuremath{p_{\mathrm{T}}}\xspace}

\newcommand{\lum} {\ensuremath{\mathcal{L}}\xspace}

\def\evtgen     {\mbox{\textsc{EvtGen}}\xspace}

\def\geant      {\mbox{\textsc{Geant4}}\xspace}

\def\photos     {\mbox{\textsc{Photos}}\xspace}

\def\pythia     {\mbox{\textsc{Pythia}}\xspace}

\def\tell1  {TELL1\xspace}
\def\ukl1   {UKL1\xspace}

\newcommand{\lhcborcid}[1]{\href{https://orcid.org/#1}{\hspace*{0.1em}\raisebox{-0.45ex}{\includegraphics[width=1em]{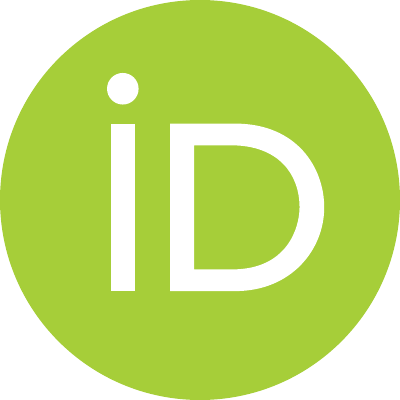}}}}


%% file: title-LHCb-PAPER.tex
\begin{titlepage}
\pagenumbering{roman}

\vspace*{-1.5cm}
\centerline{\large EUROPEAN ORGANIZATION FOR NUCLEAR RESEARCH (CERN)}
\vspace*{1.5cm}
\noindent
\begin{tabular*}{\linewidth}{lc@{\extracolsep{\fill}}r@{\extracolsep{0pt}}}
\ifthenelse{\boolean{pdflatex}}
{\vspace*{-1.5cm}\mbox{\!\!\!\includegraphics[width=.14\textwidth]{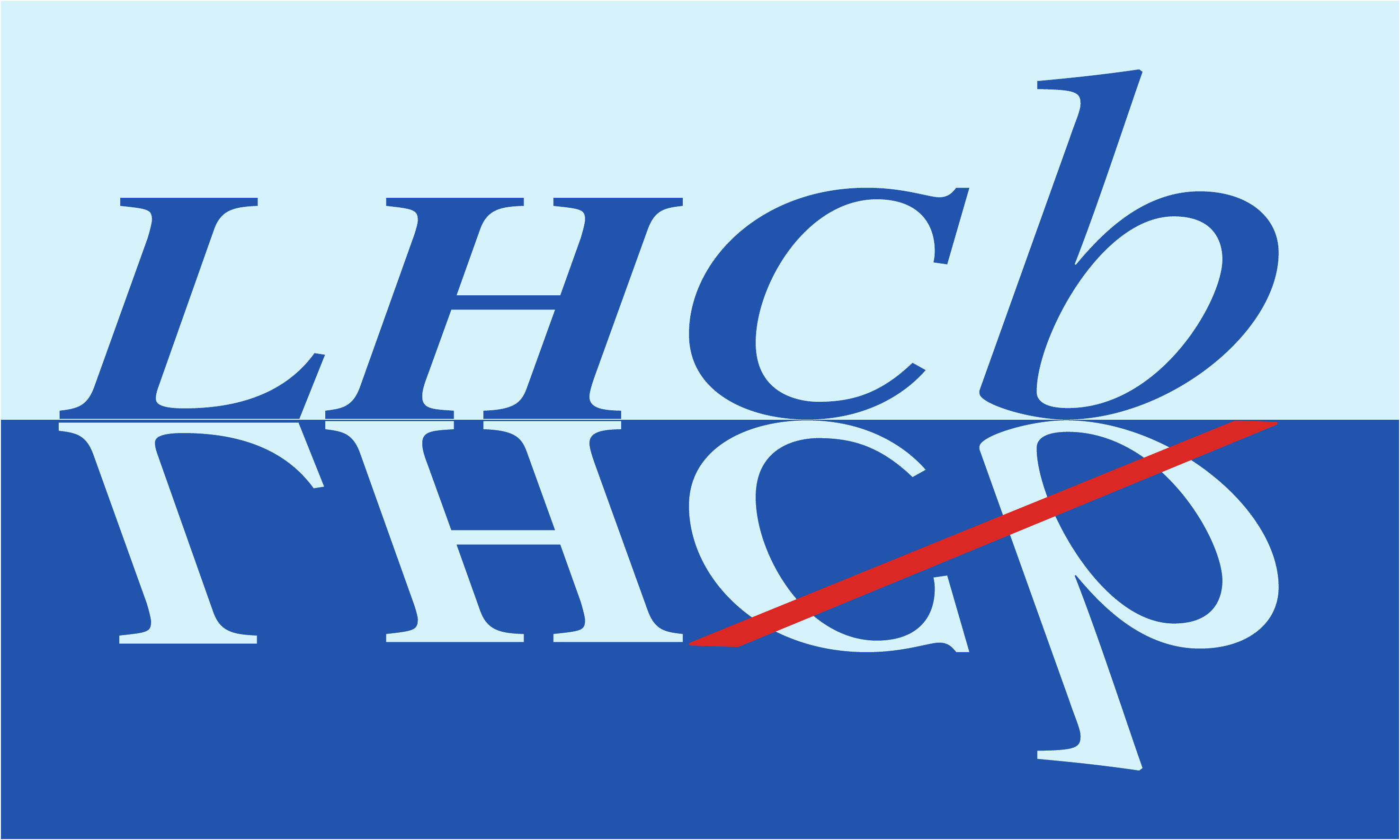}} & &}%
{\vspace*{-1.2cm}\mbox{\!\!\!\includegraphics[width=.12\textwidth]{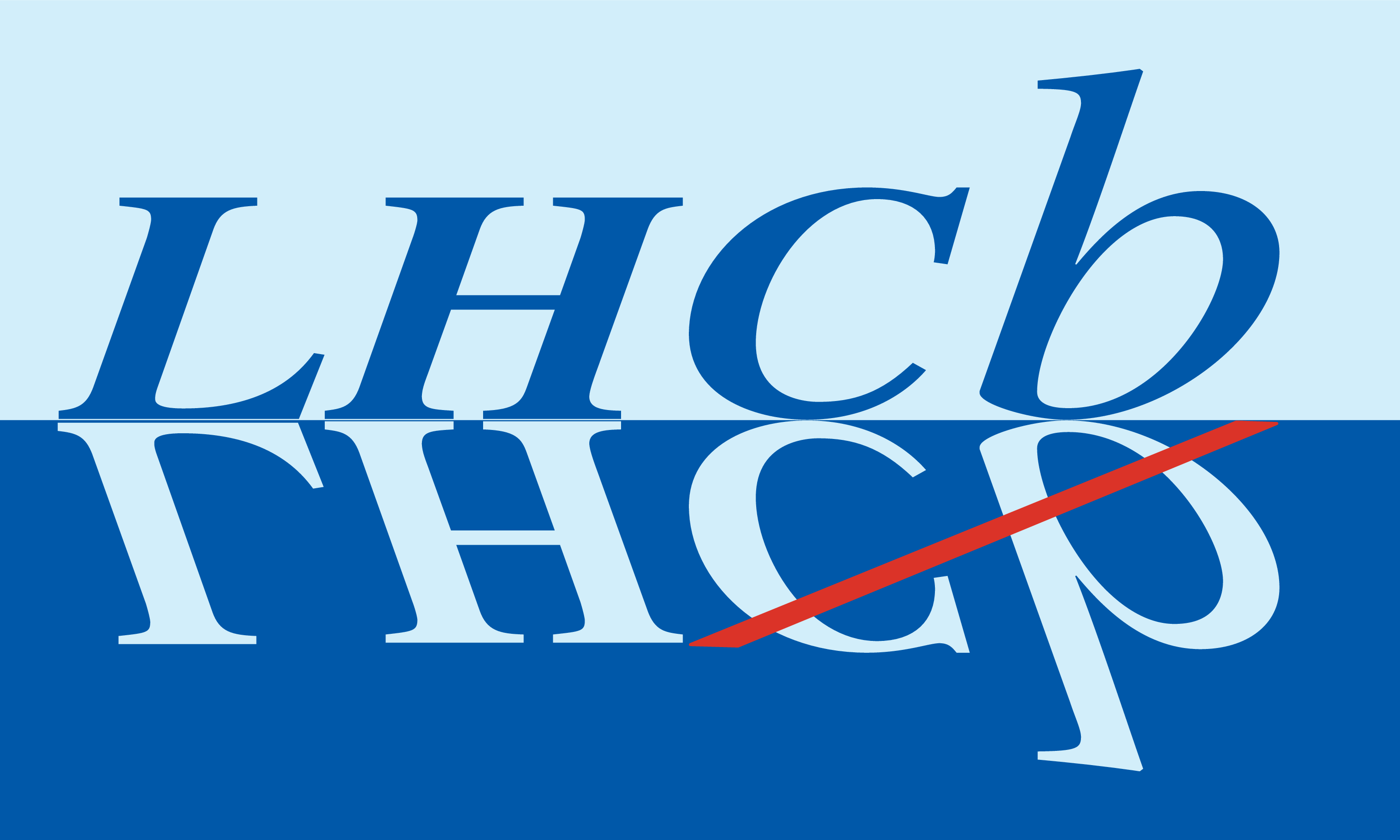}} & &}%
\\
 & & CERN-EP-2024-318 \\  
 & & LHCb-PAPER-2024-038 \\  
 & & \today \\ 
 & & \\
\end{tabular*}

\vspace*{2.0cm}

{\normalfont\bfseries\boldmath\huge
\begin{center}
  \papertitle 
\end{center}
}

\vspace*{2.0cm}

\begin{center}

\paperauthors\footnote{Authors are listed at the end of this paper.}
\end{center}

\vspace{\fill}

\begin{abstract}
  \noindent
  The $\mit{\Upsilon}(\mathrm{2}S)$ and $\mit{\Upsilon}(\mathrm{3}S)$ production cross-sections are measured relative to that of the $\mit{\Upsilon}(\mathrm{1}S)$ meson, as a function of charged-particle multiplicity in proton-proton collisions at a centre-of-mass energy of $13\tev$. The measurement uses data collected by the LHCb experiment in 2018 corresponding to an integrated luminosity of 2$\text{\,fb}^{-1}$. 
  Both the $\mit{\Upsilon}(\mathrm{2}S)$-to-$\mit{\Upsilon}(\mathrm{1}S)$ and $\mit{\Upsilon}(\mathrm{3}S)$-to-$\mit{\Upsilon}(\mathrm{1}S)$ cross-section ratios are found to decrease significantly as a function of event multiplicity, 
  with the  $\mit{\Upsilon}(\mathrm{3}S)$-to-$\mit{\Upsilon}(\mathrm{1}S)$ ratio showing a steeper decline towards high multiplicity. 
  This hierarchy is qualitatively consistent with the comover model predictions, indicating that final-state interactions play an important role in bottomonia production in high-multiplicity events.

\end{abstract}

\vspace*{2.0cm}

\begin{center}
Published in JHEP 05 (2025) 011
\end{center}

\vspace{\fill}

{\footnotesize 
\centerline{\copyright~\papercopyright. \href{\paperlicenceurl}{\paperlicence}.}}
\vspace*{2mm}

\end{titlepage}


\newpage
\setcounter{page}{2}
\mbox{~}

%% file: body.tex
\section{Introduction}
\label{sec:Introduction}

At the extremely high temperatures and densities encountered in relativistic heavy-ion collisions, the coupling constant of QCD becomes so small that quarks and gluons are no longer confined in hadrons~\cite{Aoki:2006we,HotQCD:2018pds}. 
This state of matter, called quark-gluon plasma (QGP)~\cite{Shuryak:1978ij,Gross:1980br}, exhibits special features that appear as a result of hot nuclear matter effects~\cite{Adams_2005}. 
Due to Debye colour screening, surrounding partons prevent heavy quarks from combining into quarkonia\cite{Matsui:1986dk}, resulting in the suppression of heavy-flavour quarkonia production such as the different $\PUpsilon$ states, hereafter denoted $\Yn{n}$ (where $n=1, 2, \ldots$).  Additionally, the degree of  suppression increases the more weakly bound an $\Yn{n}$ state is, 
meaning that the $\Yn{3}$ is more suppressed than the $\Yn{2}$. The screening length is related to the temperature of the QGP which can thus be probed through the study of quarkonia production~\cite{Mocsy:2007jz}.

Many crucial heavy-quarkonia production measurements in heavy-ion collisions have indeed indicated the existence of QGP~\cite{STAR:2022rpk,STAR:2019fge,ALICE:2013osk,CMS:2018zza}. 
Furthermore, measurements from $p$A (where A represents a heavy nucleus such as Au or Pb) also showed suppression~\cite{ALICE:2018szk,LHCb:2013gmv,LHCb:2018psc}. This is unexpected because the energy density of these colliding systems is too low to form the QGP. 
Therefore, another explanation has been proposed for this suppression, namely cold nuclear matter (CNM) effects, which include contributions from both initial- and final-state effects.
Initial-state effects arise from the modification of nucleon parton-distribution functions for ions~\cite{Eskola:2009uj}, energy loss~\cite{Arleo:2014oha} and multiparticle scattering~\cite{Kang:2014hha}. 
Final-state effects include interactions with comoving particles\cite{Ferreiro:2012rq,Ferreiro:2014bia}, known as the comover effect. During the collision, the comoving particles are produced in the same spatial and temporal region as those of the quarkonia. They travel together and interact with the quarkonia, which leads to the suppression of quarkonia production.
Initial-state effects influence the production of the different $\PUpsilon$ states equally. 
In measurements of the production ratios of different $\PUpsilon$ states, these initial-state effects cancel, thus isolating final-state contributions such as the comover effect.

Quarkonia suppression in proton-proton ($pp$) collisions should provide a baseline to study the final-state effects in $p$A and AA collisions. In particular,
$pp$ collisions with high charged-particle multiplicity are expected to provide an environment similar to that of $p$A and even AA collisions, thus can be used as a probe of final-state contributions. Recently, phenomena specific to AA collisions have been observed in high-multiplicity $pp$ collisions, such as collective flow~\cite{CMS:2016fnw} and strangeness enhancement~\cite{ALICE:2016fzo}.
It is therefore plausible that the final-state effects successfully describing the suppression of quarkonia in $p$A collisions can be observed in $pp$ collisions. 

The suppression of quarkonium production across different collision systems is under intense experimental scrutiny~\cite{CMS:2020fae,ALICE:2016flj,ALICE:2018wzm,ALICE:2020eji,CMS:2020fae,LHCb:2020sey,ALICE:2022yzs,PHENIX:2022nrm,ATLAS:2022exb,LHCb:2023xie}.
For example, the CMS collaboration measured the event-activity dependence of $\Yn{n}$ production cross-section ratios in  $\sqrt{s}=7\tev$ $pp$ collisions~\cite{CMS:2020fae}, finding that the ratios $\Yn{3}$/$\Yn{1}$ and $\Yn{2}$/$\Yn{1}$ decrease with track multiplicity. However, the ALICE collaboration measured the ratio $\Yn{n}$/$\Yn{1}$ as a function of
charged-particle multiplicity in $pp$ collisions\cite{ALICE:2022yzs}, and did not find a significant dependence on multiplicity. Recently, the LHCb collaboration observed that the ratio $\psitwos$ over $\jpsi$ decreases with multiplicity~\cite{LHCb:2023xie}. The measurement reported here complements these previous studies, and further explores quarkonia production and possible suppression mechanisms in the forward rapidity region. 

In this analysis, the production cross-sections of the $\Yn{2}$ and $\Yn{3}$ states relative to that of the $\Yn{1}$ state are measured using data collected in 2018 with an integrated luminosity of $2\invfb$ in $pp$ collisions at $\sqrt{s}=13\tev$. Results are obtained in three-dimensional bins of transverse momentum (\pt), \mbox{rapidity ($y$)} and multiplicity. The measurements are performed in the kinematic range $2.0 <y<4.5$ and \mbox{$0 <\pt <30 \gevc$}.
Yields of the $\Yn{n}$ states are extracted from fits to the $\mup\mun$ mass spectrum. 
To distinguish the effects from QGP and interactions with comoving particles, multiplicity is characterised by three variables: 
the total number of tracks used for reconstructing the primary vertex (PV) $\pvt$, which is a good estimator for the collision multiplicity~\cite{LHCb-DP-2014-001, Kucharczyk:1756296}, and the number of backward and forward tracks among PV tracks, referred to as $\bkt$ and $\fwt$, respectively. 
The variable $\pvt$ and $\fwt$ include tracks from the $\Yn{n}$  decay in the determination of multiplicity, while $\bkt$ is measured in the opposite direction, where the $\Yn{n}$ decay tracks are not included.

\section{Detector and dataset}
The LHCb detector~\cite{LHCb-DP-2008-001,LHCb-DP-2014-002} is a single-arm forward spectrometer covering the pseudorapidity range of $2 < \eta <5$, primarily designed for the investigation of particles containing $b$ or $c$ quarks. This detector comprises a high-precision tracking system, which includes a silicon-strip vertex detector (VELO) surrounding the $pp$ interaction region~\cite{LHCb-DP-2014-001}, a large-area silicon-strip detector located upstream of a dipole magnet with a bending power of about 4\,T\,m, and three stations of silicon-strip detectors in the high $\eta$ region and straw drift tubes in the low $\eta$ region~\cite{LHCb-DP-2013-003} placed downstream of the magnet. 
The tracking system provides the measurement of the momentum, $p$, of charged particles with a relative uncertainty ranging from $0.5\%$ below 20\gevc to $1.0\%$ around 200\gevc. The minimum distance of a track to a PV, known as the impact parameter, is determined with a resolution of $(15 + 29/\pt)\mum$, where \pt is expressed in \gevc.
Different types of charged hadrons are distinguished using information from two ring-imaging Cherenkov detectors~\cite{LHCb-DP-2012-003}. 
Photons, electrons, and hadrons are identified by a calorimeter system consisting of scintillating-pad (SPD) and preshower detectors, an electromagnetic and a hadronic calorimeter. 
Muons are identified by a system composed of alternating layers of iron and multiwire proportional chambers~\cite{LHCb-DP-2012-002}.

Candidate  $\Yn{n}$ mesons are reconstructed using $\mumu$  pairs selected by the trigger~\cite{LHCb-DP-2019-001}. 
The trigger selection consists of a hardware and a software stage. The hardware stage selects dimuon candidates with loose requirements for the occupancy in the SPD and the \pt of the candidate. 
The software stage requires two oppositely charged muon candidates with large momenta and transverse momenta. 
The invariant mass of the muon pair must be greater than 7900\,\mevcc. 
Particle identification (PID) requirements are used to reduce the background from hadrons reaching the muon system and being wrongly identified as muons. To suppress the contribution from other interactions to the measured multiplicity, 
only events with a single visible $pp$ interaction per bunch-crossing are selected.

The PV coordinate along the beam direction, $z_\mathrm{PV}$, is restricted to the region where the VELO acceptance is uniform. The allowed $z_\mathrm{PV}$ ranges are listed in Table~\ref{MulScheme}. 
An unbiased data sample recorded by a random trigger, called \mbox{NoBias} data, is used to obtain a reference distribution for the multiplicity. By normalising with the mean value in \mbox{NoBias} data, the multiplicity can be compared across different collision systems~\cite{KOBA1972317}.


To reduce combinatorial background, additional selections are applied to the $\Yn{n}$ candidates. 
First, the muons are required to form a vertex and to pass more stringent PID requirements than those imposed at the trigger level. 
The momentum of each muon is required to be more than $10\gevc$, and $\pt$ to be greater than $1\gevc$. 
Furthermore, the muons are required to be within the pseudorapidity range \mbox{$1.9 < \eta < 4.9$} and the dimuon invariant mass must be within \mbox{$8800< m_{\mumu} < 10700\mevcc$}. Combined with the track momentum resolution performance, these criteria allow a clear separation of the 
different $\Yn{n}$ states.
\begin{table}[t]
    \centering
    \caption{The binning schemes and $z_\mathrm{PV}$ ranges for different multiplicity variables. The binning schemes are presented in terms of the bin edges in multiplicity.
}
    \begin{tabular}{ccc}
    \hline
    Variable & Binning schemes & $z_\mathrm{PV} $\ ranges\\
    \hline
        $\pvt$ & [0, 35, 48, 60, 73, 90, 200]  & [\phantom{1}$-50$, 160]~mm \\
        $\fwt$ & [0, 22, 32, 40, 49, 61, 151]  & [$-160$, 160]~mm\\
        $\bkt$ & [0, \phantom{0}8, 13, 18, 23, 31, \phantom{0}91] & [\phantom{1}$-20$, 160]~mm\\
    \hline
    \end{tabular}
    \label{MulScheme}
\end{table}

Simulation samples are produced to model the efficiencies and the mass distribution of the signal candidates. 
 The \pythia~\cite{Sjostrand:2006za} generator with a specific LHCb configuration~\cite{LHCb-PROC-2010-056} is used to generate $pp$ collisions. Bottomonia decays
are described by \evtgen~\cite{Lange:2001uf}, in which final-state radiation is generated using \photos~\cite{Golonka:2005pn}. The interaction of the generated particles with the detector and its
response are implemented using the \geant
toolkit~\cite{Allison:2006ve, *Agostinelli:2002hh} as described in Ref.~\cite{LHCb-PROC-2011-006}. 
The bottomonia states are generated unpolarised in accordance with a recent measurement~\cite{LHCb:2017scf}, where the polarisation is found to be close to zero. The effects of possible nonzero polarisation are considered as a source of systematic uncertainties as described in Section~\ref{sec:systematics}.
\section{Determination of the production ratios}
\label{sec:xsec}
The double-differential production cross-section is defined as
\begin{equation}
\label{eq:xsec}
  \frac{\deriv^2\sigma_{\Yn{n}}}{\deriv y\, \deriv \pt}
  = \frac{N(\Yn{n};\pt,y)}
         {\lum\times\etot(\pt,y)\times\BR(\Yn{n}\to\mumu)\times\Delta \pt \times \Delta y}, \quad(n=1,2,3),
\end{equation}
where $N(\Yn{n};\pt,y)$ is the signal yield extracted by fitting the dimuon invariant-mass distribution in each $(\pt,y)$ bin, \lum is the integrated luminosity, $\etot(\pt,y)$ is the total efficiency calculated using simulation samples, and $\Delta \pt$ and $\Delta y$ are the bin widths for transverse momentum and rapidity, respectively. 
The branching fraction \mbox{$\BR(\Yn{n}\to\mumu)$} is $(2.48\pm 0.04) \%$ for $\Yn{1}$, $(1.93\pm 0.17) \%$ for $\Yn{2}$ and $(2.18 \pm 0.21) \%$ for $\Yn{3}$ states~\cite{PDG2024}. 
According to Eq.~\ref{eq:xsec}, the double-differential production ratio is defined as
\begin{equation}
  \frac{\sigma_{\Yn{n}}}{\sigma_{\Yn{1}}} (\pt, y)
  = \frac{N(\Yn{n};\pt,y)}{N(\Yn{1};\pt,y)}\times\frac{\etot(\Yn{1};\pt,y)}{\etot(\Yn{n};\pt,y)}\times
        \frac {\BR(\Yn{1}\to\mumu)}{\BR(\Yn{n}\to\mumu)}, \quad(n=2,3).
\end{equation}

The measurements are performed in different multiplicity bins according to the binning schemes shown in Table~\ref{MulScheme}. 
\begin{table}[t]
    \centering
    \renewcommand*{\arraystretch}{1.1}
    \caption{The mean value of different multiplicity variables in NoBias data.}
\begin{tabular}{cc}
\hline
    Variables & Mean value \\
    \hline
        $\pvt$ & 26.03  \\
        $\fwt$& 16.17 \\
        $\bkt$& \;\,9.86 \\
    \hline
    \end{tabular}
   \label{MulNobias}
\end{table} 
The binning schemes are chosen so that there are sufficient signal candidates to reliably extract the signal yields within each multiplicity bin.
The normalised multiplicity is defined as
\begin{equation}
\label{eq:self_mul}
   \langle M \rangle_{i} / \langle M \rangle_\text{NoBias}
 = \frac{\sum_{n=1}^{3} \omega_{i,n}\langle M\rangle_{i,n}}{\sum_{n=1}^{3} \omega_{i,n}} \times \frac{1}{\langle M \rangle_\text{NoBias}},
\end{equation}
where $M$ is the multiplicity variable ($\pvt$, $\bkt$ or $\fwt$), $i$ is the index for the multiplicity bin, $\omega_{i,n}$ is the inverse variance of $\langle M \rangle_{i,n}$, and $\langle M \rangle_\text{NoBias}$ is the average multiplicity in \mbox{NoBias} data. The invariant mass is used as a discriminating variable to extract the multiplicity distribution for different $\Yn{n}$ states with the \sPlot method~\cite{Pivk:2004ty}.
The multiplicity variables are self-normalised according to their corresponding mean values in \mbox{NoBias} data according to Eq.~\ref{eq:self_mul}, so that they can be compared across different collision systems under different centre-of-mass energies~\cite{KOBA1972317}. 
The mean values in \mbox{NoBias} data can be found in Table~\ref{MulNobias}.

The production ratio over a specific kinematic region can be obtained by
\begin{equation}
\label{eq:intsxec}
\frac{\sigma_{\Yn{n}}}{\sigma_{\Yn{1}}}
 = \frac{\sum_{k\in \mathbf{K}}N(\Yn{n})_k/\etot(\Yn{n})_k}{\sum_{k\in \mathbf{K}}N(\Yn{1})_k/\etot(\Yn{1})_k} \times  \frac {\BR(\Yn{1}\to\mumu)}{\BR(\Yn{n}\to\mumu)}, \quad(n=2,3),
\end{equation}
where $\mathbf{K}$ is a certain set of $(\pt,y)$ bins. Finally, the production ratio in bins of multiplicity is normalised by the total production ratio,
\begin{small}
\begin{equation}
\begin{aligned}
\text{Normalised} \ \frac{\sigma_{\Yn{n},i}}{\sigma_{\Yn{1},i}}
=\frac{N(\Yn{n})_i/\etot(\Yn{n})_i}{\sum_{i\in \mathbf{I}}N(\Yn{n})_i/\etot(\Yn{n})_i}\times\frac{\sum_{i\in \mathbf{I}}N(\Yn{1})_i/\etot(\Yn{1})_i}{N(\Yn{1})_i/\etot(\Yn{1})_i},\\
 \quad(n=2,3),
\end{aligned}
\end{equation}
\end{small}

\noindent where $\mathbf{I}$ is the set of multiplicity bins, and $i$ is the index for multiplicity bins. By comparing the normalised ratio to unity, the multiplicity dependence can be compared for different states. 

\begin{figure}[!tbp]
\centering
\begin{minipage}[t]{0.7\textwidth}
\centering
\includegraphics[width=1.0\textwidth]{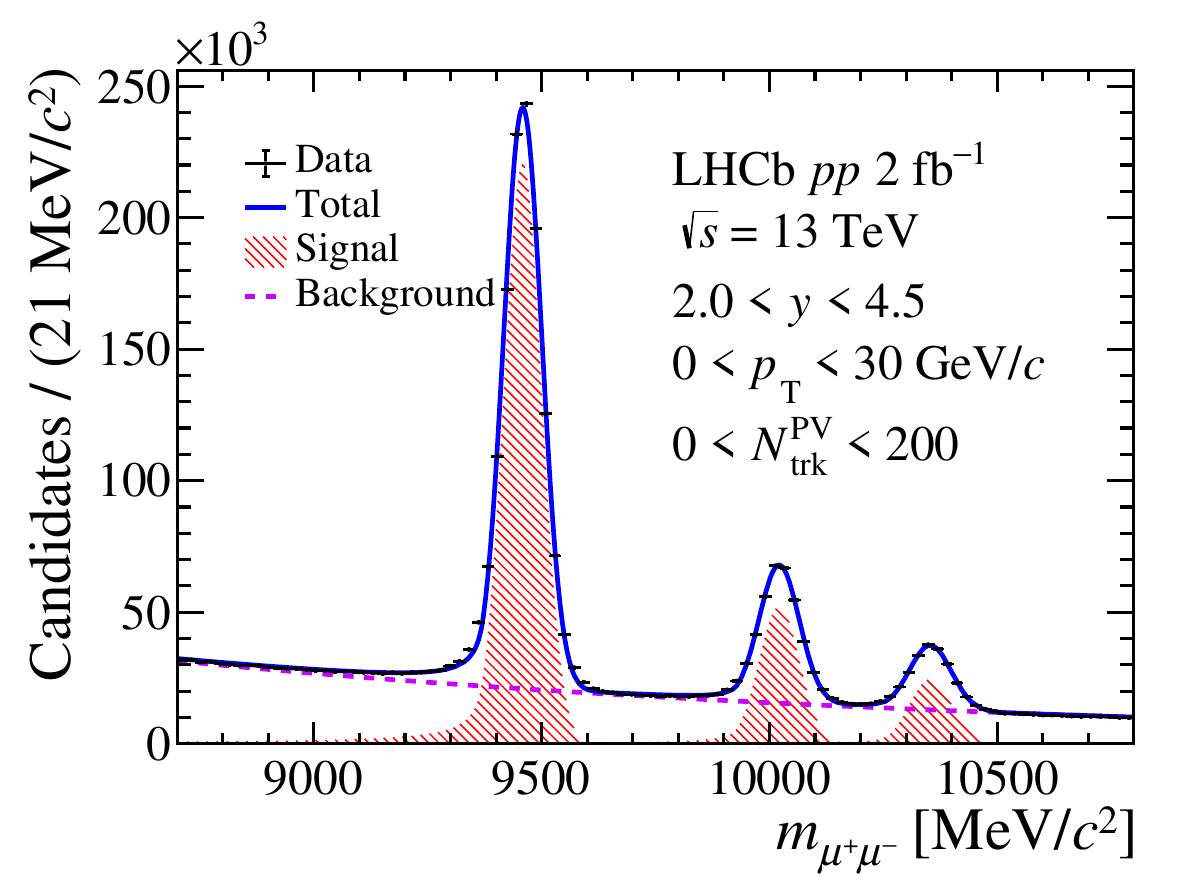}
\end{minipage}
\caption{Invariant-mass distribution of 
$\mu^+\mu^-$ candidates. The fit results are also shown.}
\label{fig:dimuonmass}
\end{figure}

The yields are extracted by performing an extended unbinned maximum-likelihood fit to the dimuon invariant-mass distributions. The signal of the mass spectrum is described by a Crystal Ball function~\cite{Skwarnicki:1986xj} and the background is described by an exponential function. To reduce the number of free parameters of the model, the relation between the width and tail parameters of the Crystal Ball function is parametrised from simulation. The mass differences between $\Yn{n}$ states are fixed to the current world averages\cite{PDG2024}. The fit to the full sample is shown in Fig.~\ref{fig:dimuonmass}. The fit results in each bin have good qualities and are guaranteed by the chi-square per degree of freedom, $\chisqndf$. 

The total efficiency is a product of the detector's geometrical acceptance and efficiencies of the particle reconstruction and selection, muon identification, and trigger. 
All efficiencies are calculated using simulation samples and corrected by the data-driven methods described in Refs.~\cite{Anderlini:2202412,LHCb-DP-2013-002}. 
The efficiencies are determined in $(\pt,y, \text{multiplicity})$ bins according to the binning scheme used for data. They are found to be similar for the different \Yn{n} states considered.

\section{Systematic uncertainties}
\label{sec:systematics}
The considered sources of systematic uncertainties are associated with the determination of signal yields and evaluation of the efficiencies. 
Systematic uncertainties for the invariant-mass fit are 
considered separately for the signal and background shapes in the whole kinematic region of $2.0 < y < 4.5$ and $0 
< \pt < 30 \gevc$, due to the limited sample size, and in each multiplicity bin. The systematic uncertainty for 
the signal shapes is studied by varying the shape parameters in the fit model.  The systematic uncertainty for 
the background shapes is estimated by replacing the default exponential function with a third-order polynomial 
function. In both cases, the largest relative differences between the baseline fit and the newly calculated 
production ratio are taken as systematic uncertainties.

The systematic uncertainty in the trigger efficiency calculation originates from the imperfect description of hardware and software triggers in simulation. 
The uncertainties associated with the muon and dimuon hardware triggers are studied separately due to the limited sample size. The former is studied using the tag-and-probe method~\cite{LHCb-DP-2014-002} and the latter follows the analysis in Ref.~\cite{LHCb-PAPER-2018-002}, studied with events
triggered by the single-muon hardware trigger. The uncertainty caused by software trigger requirements is estimated by data-driven methods~\cite{LHCb-DP-2012-004}, using a subset of events that are triggered independently of signal candidates.
Their quadratic sum is taken as the overall systematic uncertainty for the trigger efficiency. 

Tracking efficiencies in the simulation are calibrated using a $\jpsi\rightarrow\mup\mun$ control sample~\cite{LHCb-DP-2013-002}. An uncertainty of $0.8\%$ per track is assigned, which cancels when calculating the production ratios. The systematic uncertainty due to the limited control sample size is evaluated by the sampling method described in Ref.~\cite{LHCb:2023xie}.

The muon-identification efficiency determined using simulation is calibrated with $\jpsi\to\mumu$ decays, where the single-muon identification efficiency is measured in bins of $(p,\eta, N_{\rm{SPD}})$, where $N_{\rm{SPD}}$ is the number of SPD hits.
The uncertainty due to the limited calibration data sample size is estimated in the same way as for the tracking efficiency. 
The uncertainty related to the binning scheme of $(p,\eta, N_{\rm{SPD}})$ is studied by changing the bin sizes.

The $\Yn{n}$ states are unpolarised in the simulation, as suggested by current measurements~\cite{LHCb:2017scf,CMS:2013qur}. 
However, the uncertainty of these measurements is non-negligible and the effect of possible nonzero $\Yn{n}$ polarisation on the efficiency determination needs to be considered. 
To evaluate these effects, the simulation sample is weighted using the distribution of $\theta^*$, which is the angle between the $\mup$ in the $\Yn{n}$ rest frame and the $\Yn{n}$ momentum direction in the laboratory frame. The distribution found in the polarisation analysis is
\begin{equation}
\frac{\deriv N}{\deriv \cos \theta^*}=\frac{1+\lambda_\theta \cos ^2 \theta^*}{2+2 \times \lambda_\theta / 3},
\end{equation}
where $\lambda_\theta$ is assumed to be 0.2 according to the LHCb measurement~\cite{LHCb:2017scf}. 
The total efficiencies are calculated by weighting the $\theta^*$ 
distribution, with the difference in calculated production ratios between the weighted and baseline 
efficiencies taken as the systematic uncertainty.

The systematic uncertainties due to the limited size of the simulation sample are studied by propagating the uncertainties on the total efficiencies to the production ratio. The summary of all systematic uncertainties on the $\Yn{n}$ production ratios is given in Table~\ref{tab:SystematicSummary}. 

\begin{table}[t]
\centering
\caption{Range of systematic uncertainties (in \%) for the double-differential production ratio across different kinematic and multiplicity bins.}
\begin{center}
\begin{tabular}{cccc}
\hline
Sources& $R_{{\Yn{2})}/{\Yn{1}}}$& $R_{{\Yn{3}}/{\Yn{1}}}$&Comment\\
\hline Signal shape & $0.1-0.8$ & $0.3-1.7$ & \multirow{6}{*}{Correlated between bins} \\
 Background shape & $0.1-0.7$ & $0.0-2.3$&  \\
 Trigger efficiency & $0.3-1.5$ & $1.3-5.0$ &  \\
Track reconstruction & $0.0-0.1$& $0.0-0.1$ &\\
Muon identification & $0.0-0.8$  & $0.0-0.7$ &\\
Polarisation & $0.0-0.7$ & $0.0-0.6$  &\\
 \hline Limited simulation sample size
 & $0.9-4.6$ & $0.9-4.6$ & Bin independent\\
\hline  Total
 & $1.4-4.8$ & $1.7-5.8$ & Correlated between bins\\
\hline
\end{tabular}
\end{center}
\label{tab:SystematicSummary}
\end{table}

\section{Results}
\subsection{Multiplicity dependence of production ratios}
\begin{figure}[!tbp]
\centering
\begin{minipage}[t]{0.7\textwidth}
\centering
\includegraphics[width=1.0\textwidth]{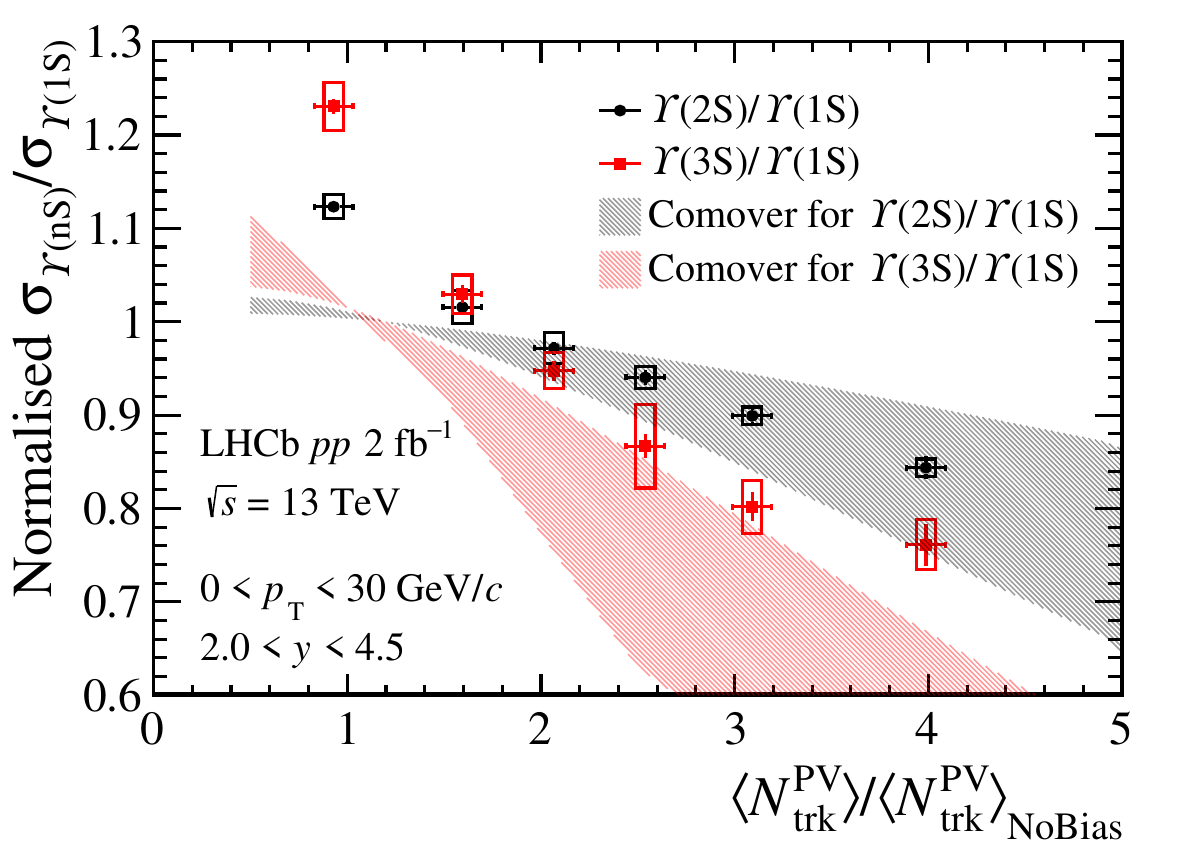}
\end{minipage}
\caption{Normalised production cross-section ratios $\mit\Upsilon(\mathrm{2}S)/\mit\Upsilon(\mathrm{1}S)$ and $\mit\Upsilon(\mathrm{3}S)/\mit\Upsilon(\mathrm{1}S)$ as a function of self-normalised $N_{\mathrm{trk}}^{\mathrm{PV}}$ for $2.0<y<4.5$ and $0<\pt<30 \text{\,Ge\kern -0.1em V\!}/c$ . The vertical error bars represent the statistical uncertainties and the boxes represent the systematic uncertainties. The width of the boxes has no physical meaning.} 
\label{fig:ratiot}
\end{figure}

The multiplicity dependence of the normalised $\Yn{n}/\Yn{1}$ production ratios with $\pvt$ over the whole kinematic region $0<\pt<30\gevc$, $2<y<4.5$ is shown in Fig.~\ref{fig:ratiot}.
Both $\Yn{2}/\Yn{1}$ and $\Yn{3}/\Yn{1}$ ratios decrease with multiplicity, and the $\Yn{3}$ state is more suppressed than the $\Yn{2}$ state. 
This hierarchical pattern is consistent with that found in 8.16\,\tev $p$Pb collisions~\cite{LHCb:2018psc}, which suggests that final-state effects such as the comover effect need to be considered in high-multiplicity $pp$ events. 
The multiplicity dependence of the $\Yn{2}/\Yn{1}$ and $\Yn{3}/\Yn{1}$ ratios with $\pvt$ are compared with comover model predictions~\cite{Ferreiro:2012rq}, which also reproduce the observed suppression pattern for $\PUpsilon$ states.
Similar to what was found for the $\psi(2S)/\jpsi$ ratio in $pp$ collisions at 13~TeV~\cite{LHCb:2023xie}, the $\Yn{2}/\Yn{1}$ ratio is in agreement with the comover model which reproduces the observed multiplicity dependence, except in the low multiplicity region. In the case of the $\Yn{3}/\Yn{1}$ ratio, the slope of the ratio is in agreement with the model but the value is in worse agreement, which motivates further theoretical investigation.

\begin{figure}[!tbp]
\centering
\begin{minipage}[t]{0.49\textwidth}
\centering
\includegraphics[width=1.0\textwidth]{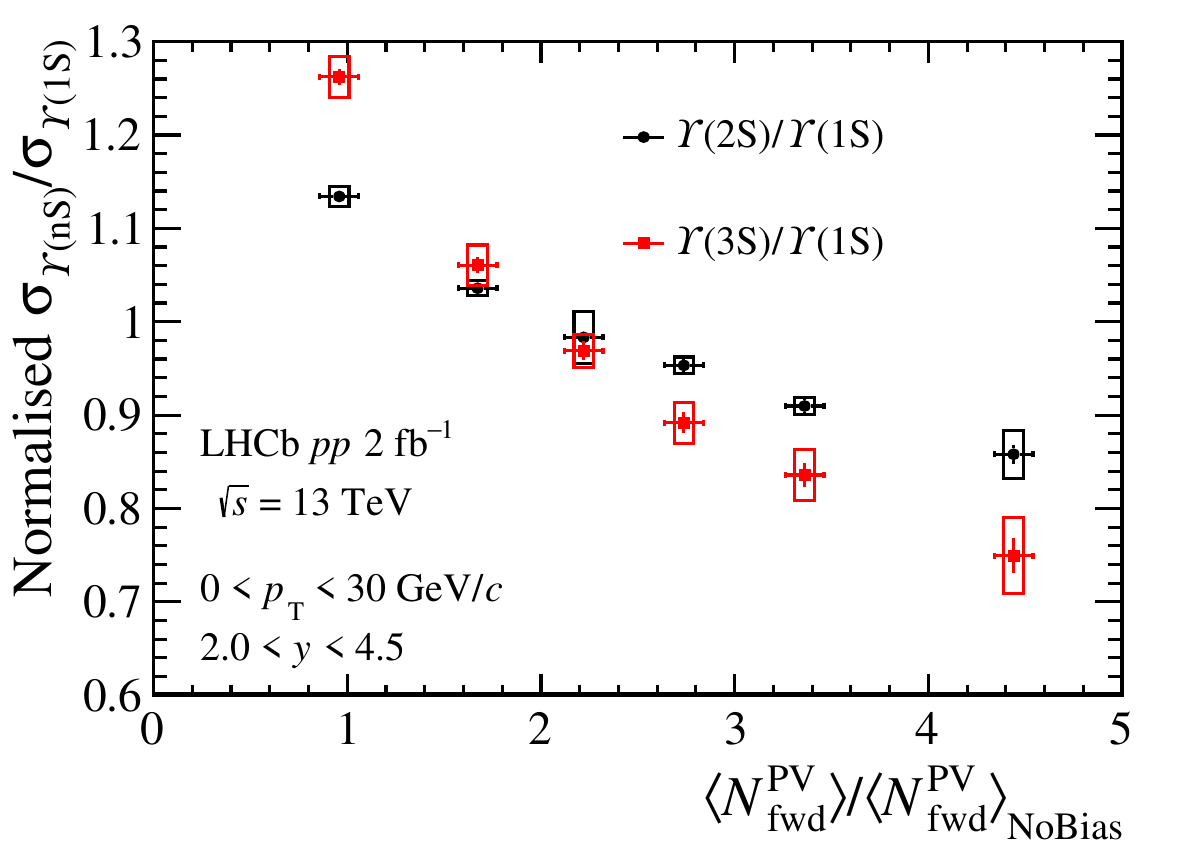}
\end{minipage}
\begin{minipage}[t]{0.49\textwidth}
\centering
\includegraphics[width=1.0\textwidth]{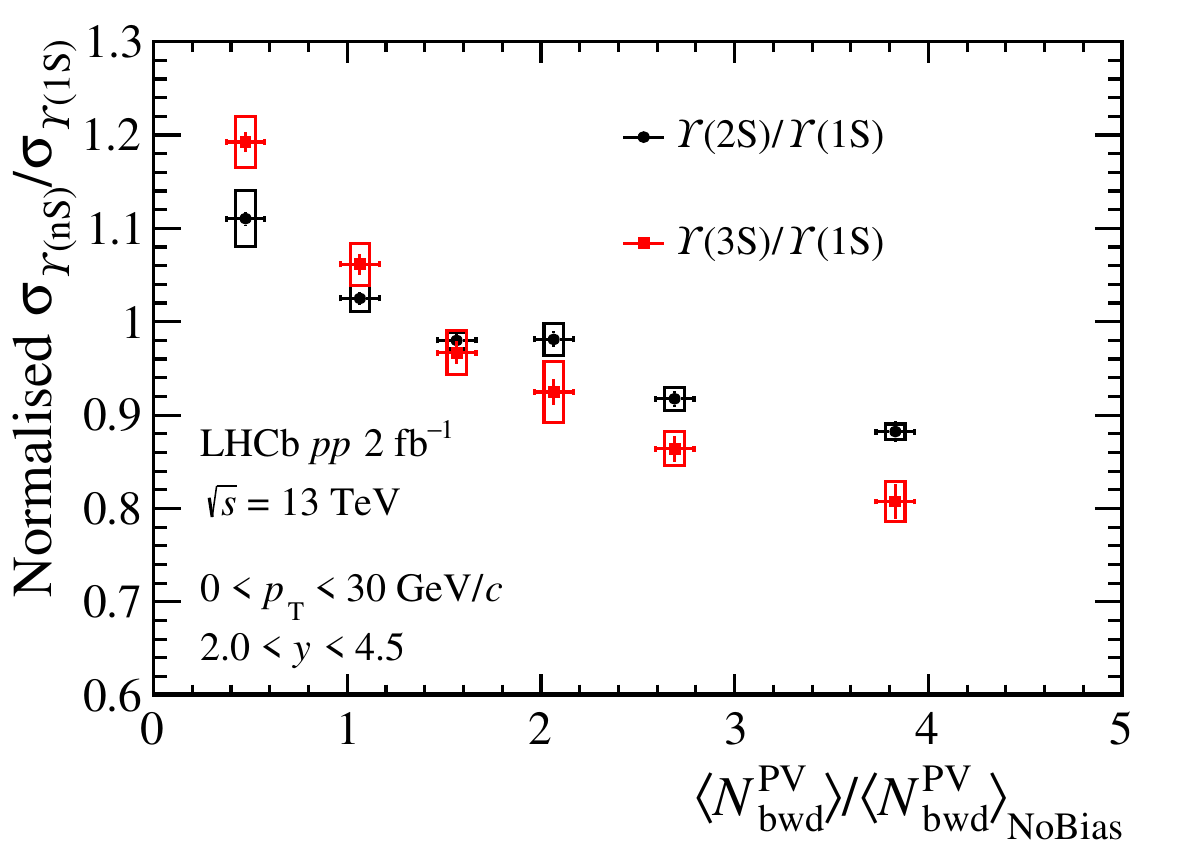}
\end{minipage}
\caption{Normalised production cross-section ratios $\mit\Upsilon(\mathrm{2}S)/\mit\Upsilon(\mathrm{1}S)$ and $\mit\Upsilon(\mathrm{3}S)/\mit\Upsilon(\mathrm{1}S)$ as a function of self-normalised (left)~$N_{\mathrm{fwd}}^{\mathrm{PV}}$ and (right)~$N_{\mathrm{bwd}}^{\mathrm{PV}}$ for $2.0<y<4.5$ and $0<\pt<30 \text{\,Ge\kern -0.1em V\!}/c$.} 
\label{fig:ratiofb}
\end{figure}

The multiplicity dependence for $\Yn{n}/\Yn{1}$ production ratios is also studied as a function of the self-normalised $\fwt$ and $\bkt$ parameters, as shown in Fig.~\ref{fig:ratiofb}. In both cases, $\Yn{3}$ production is found to be more suppressed than $\Yn{2}$ production. The $\Yn{3}/\Yn{1}$ production ratio decreases more rapidly with forward multiplicity $\fwt$ than backward multiplicity $\bkt$ since comoving particles are expected to interact with the produced bottomonia in the same rapidity range. This is consistent with the $\psi(2S)/\jpsi$ results~\cite{LHCb:2023xie}. The $\Yn{2}/\Yn{1}$ production ratios do not show a significantly steeper trend in forward compared to backward multiplicity, where the normalised cross-section ratios are from 1.13 to 0.88 within uncertainties. Though the final-state particles produced in the backward region do not interact with bottomonia, the $\Yn{n}/\Yn{1}$ ratios still decrease with $\bkt$, which results from the strong correlation between $\bkt$ and $\pvt$.

\subsection{Production ratios in different kinematic regions}
\begin{figure}[!tbp]
\includegraphics[width=0.49\textwidth]{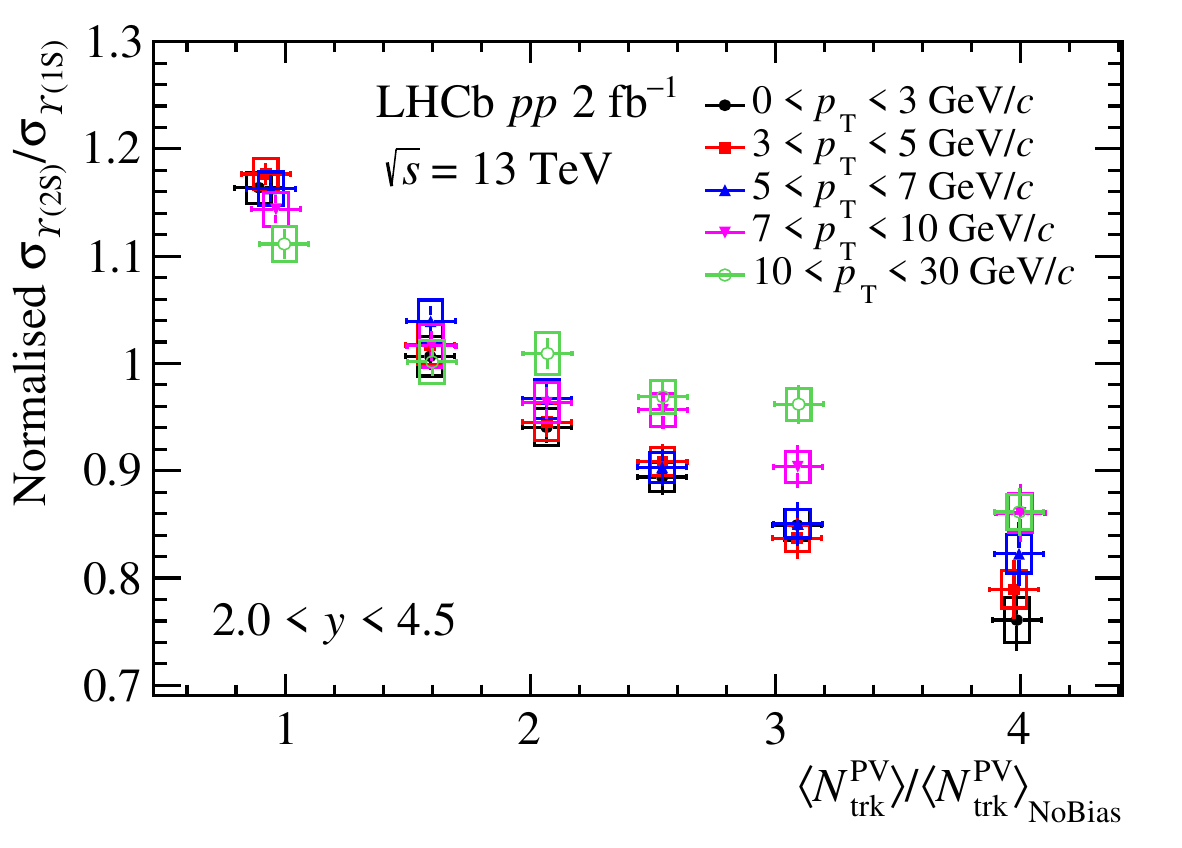}
\includegraphics[width=0.49\textwidth]{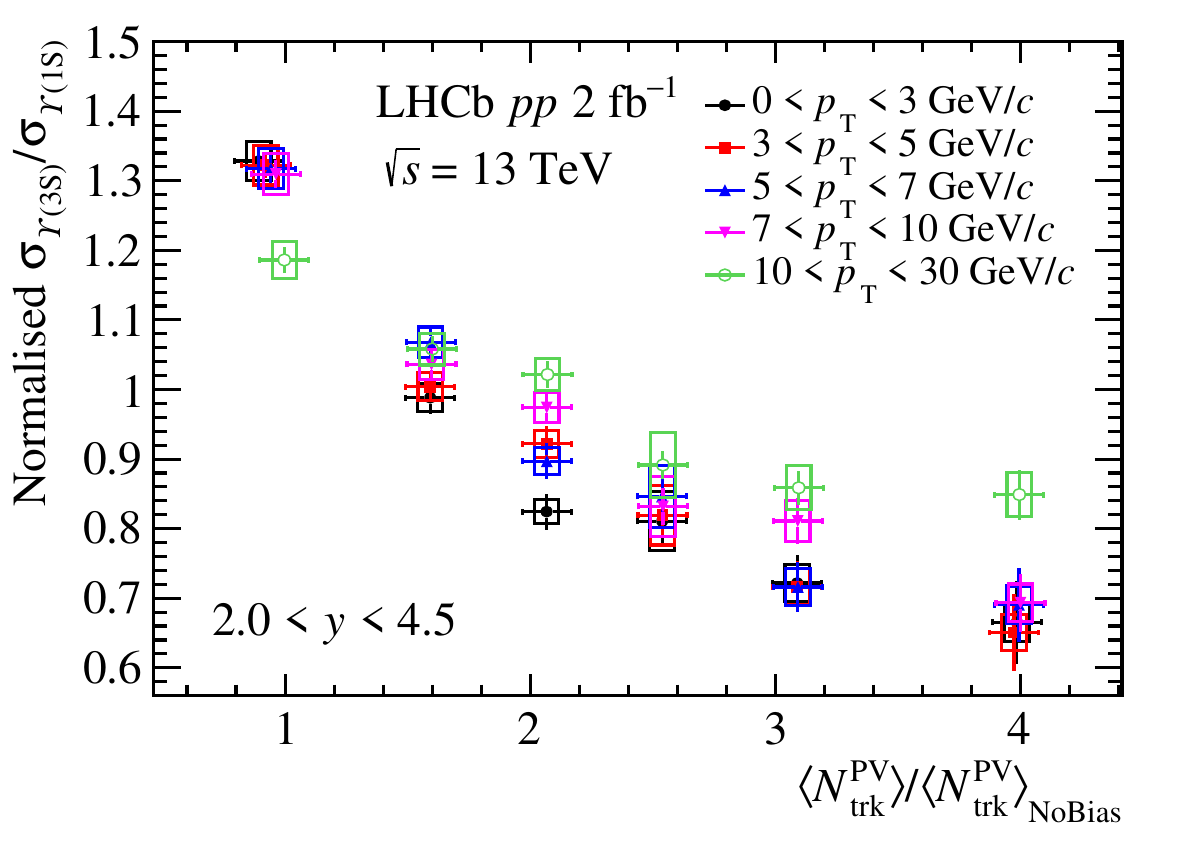}
\includegraphics[width=0.49\textwidth]{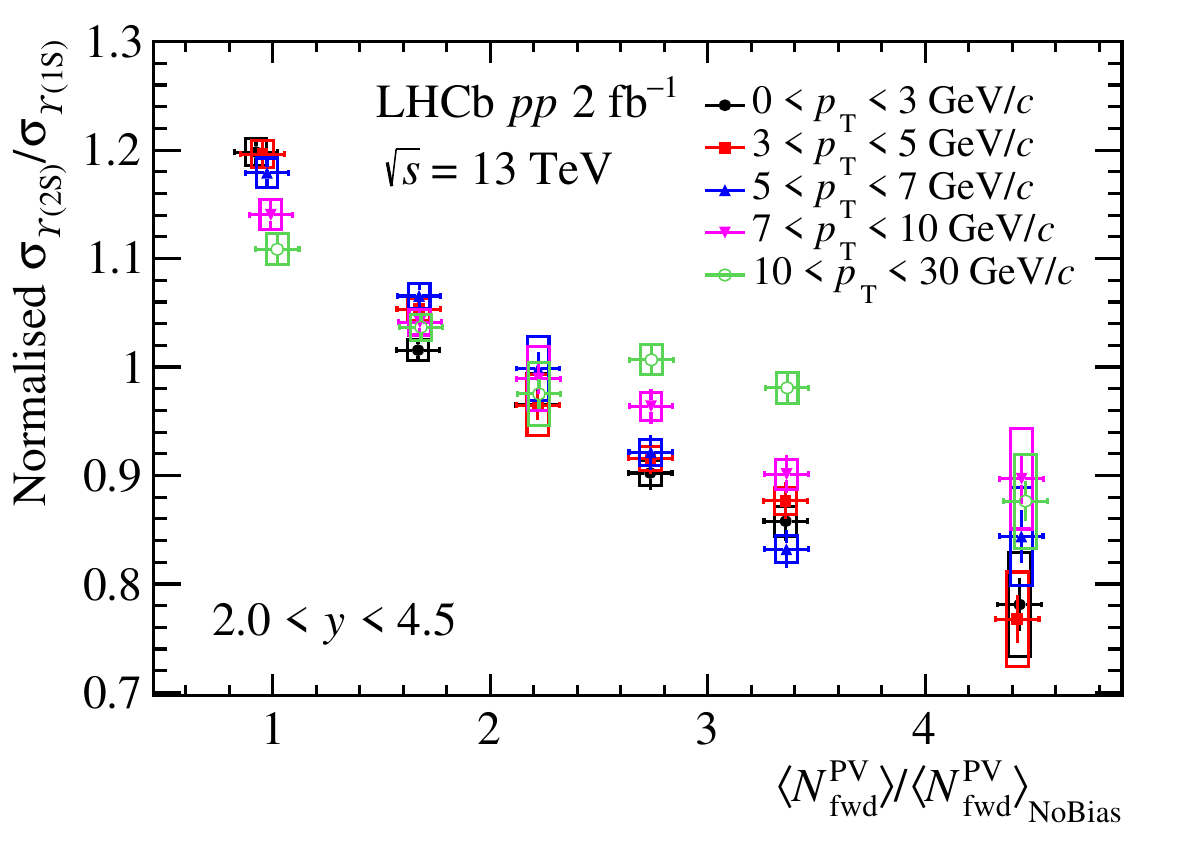}
\includegraphics[width=0.49\textwidth]{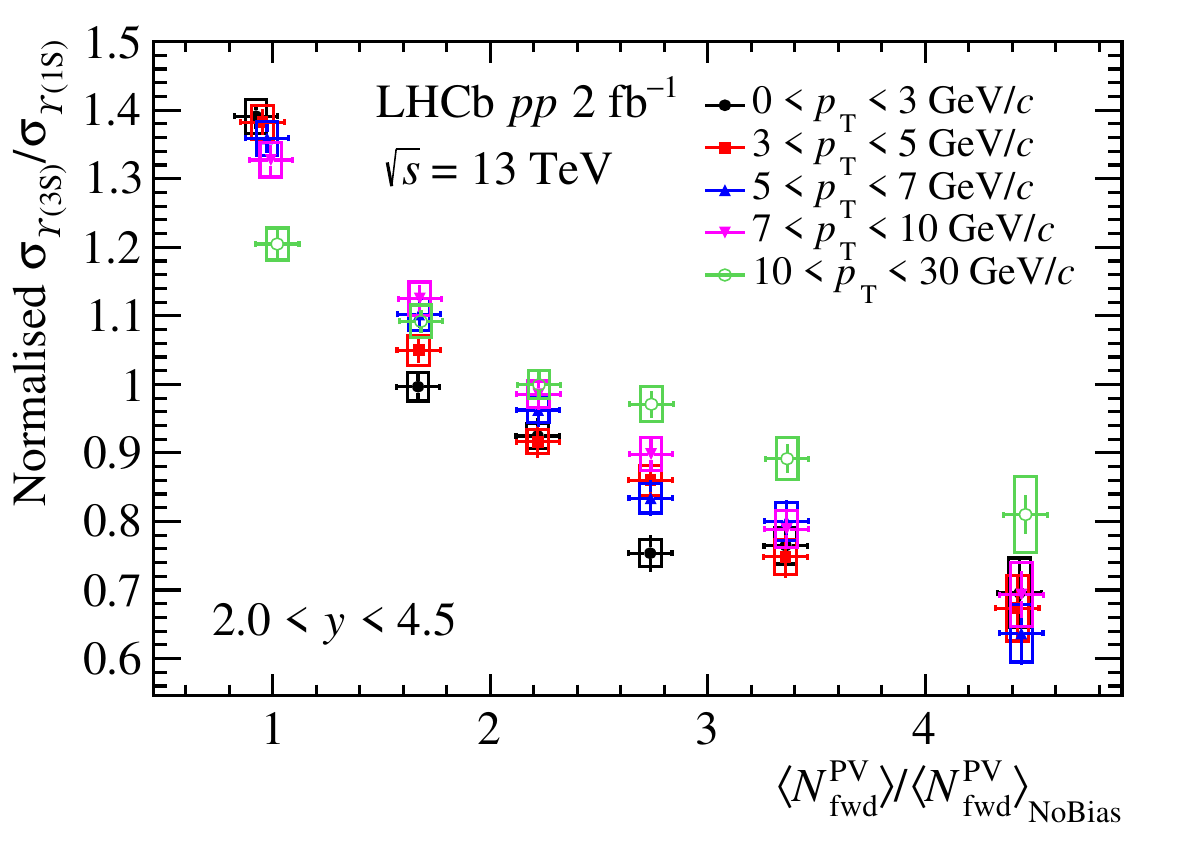}
\includegraphics[width=0.49\textwidth]{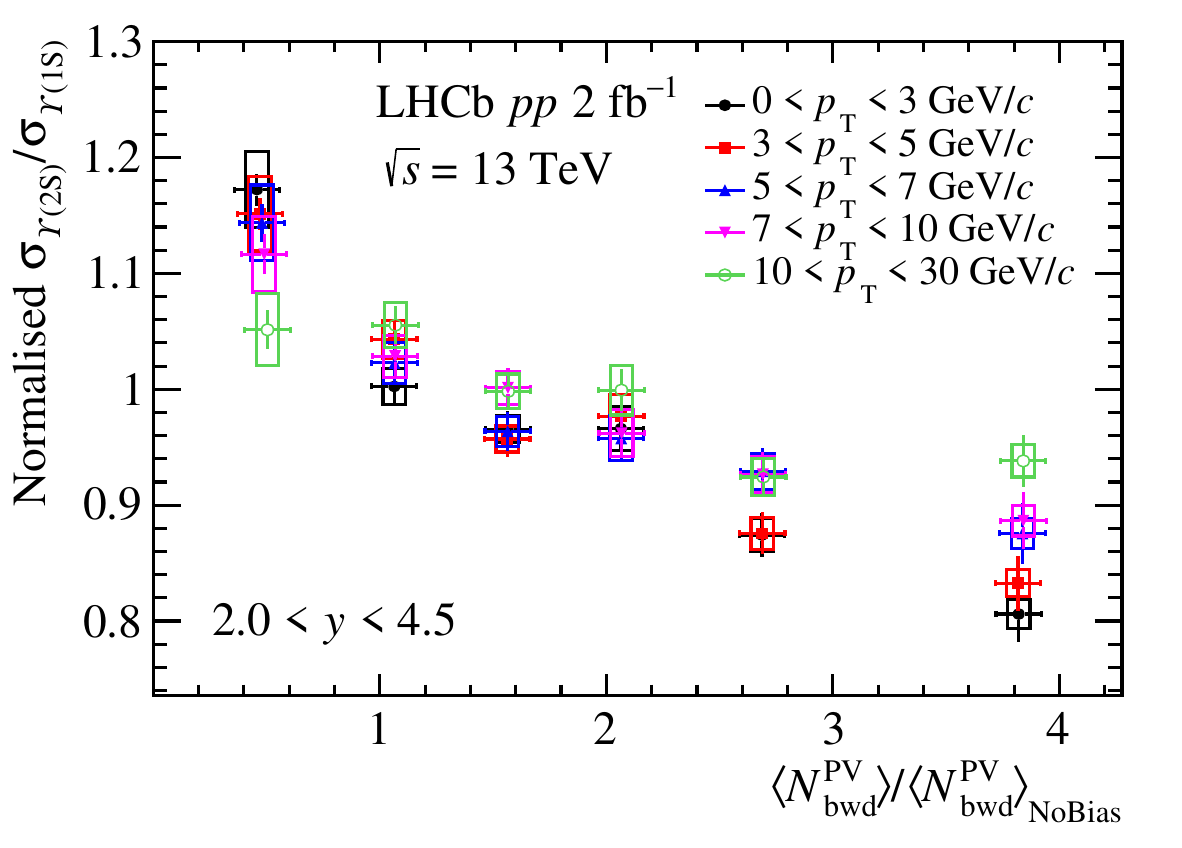}
\includegraphics[width=0.49\textwidth]{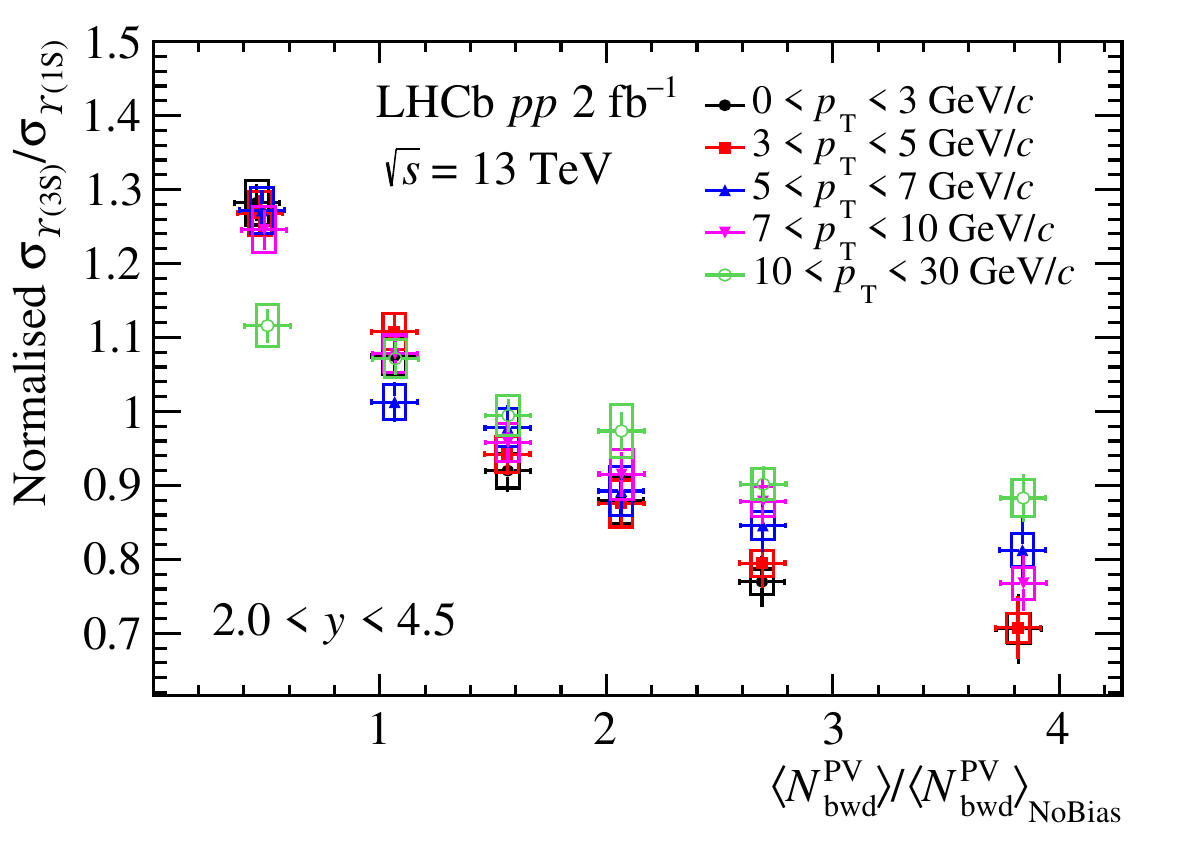}
\caption{Normalised production cross-section ratios (left)~$\mit\Upsilon(\mathrm{2}S)/\mit\Upsilon(\mathrm{1}S)$ and (right)~$\mit\Upsilon(\mathrm{3}S)/\mit\Upsilon(\mathrm{1}S)$ as a function of self-normalised (top)~$N_{\mathrm{trk}}^{\mathrm{PV}}$, (middle)~$N_{\mathrm{fwd}}^{\mathrm{PV}}$ and (bottom)~$N_{\mathrm{bwd}}^{\mathrm{PV}}$ in different $\pt$ regions for $2.0<y<4.5$.} 
\label{fig:ratioy}
\end{figure}

The multiplicity dependence of the $\Yn{n}/\Yn{1}$ production ratio in different \pt regions is shown in Fig.~\ref{fig:ratioy}. A decreasing trend is observed for the ratio of production with increasing multiplicity in each \pt bin. 
The comover model predicts an average \pt for the underlying charged particles of around 1\gevc, and thus favours stronger suppression at low \pt. 
The comover effects would then be stronger with particles emitted close to the bottomonium. In the high-\pt region, the multiplicity dependence for $\Yn{n}/\Yn{1}$ ratios declines slightly. This is consistent with CMS results in 7~TeV $pp$ collisions~\cite{CMS:2020fae} and ATLAS results in 13~TeV $pp$ collisions~\cite{ATLAS:2022xar}. 

\begin{figure}[!tbp]
\centering
\includegraphics[width=1.0\textwidth]{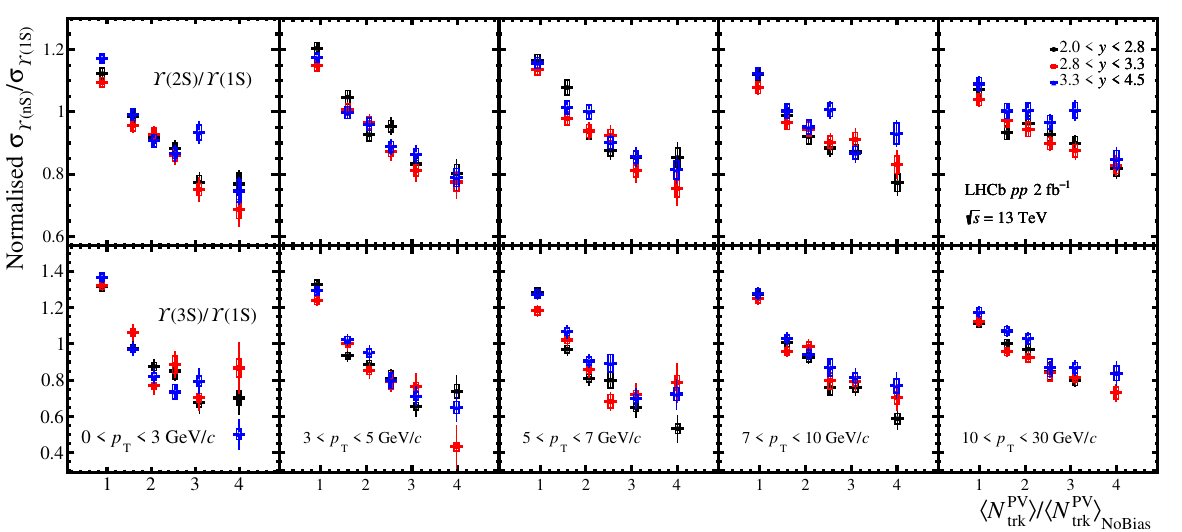}
\includegraphics[width=1.0\textwidth]{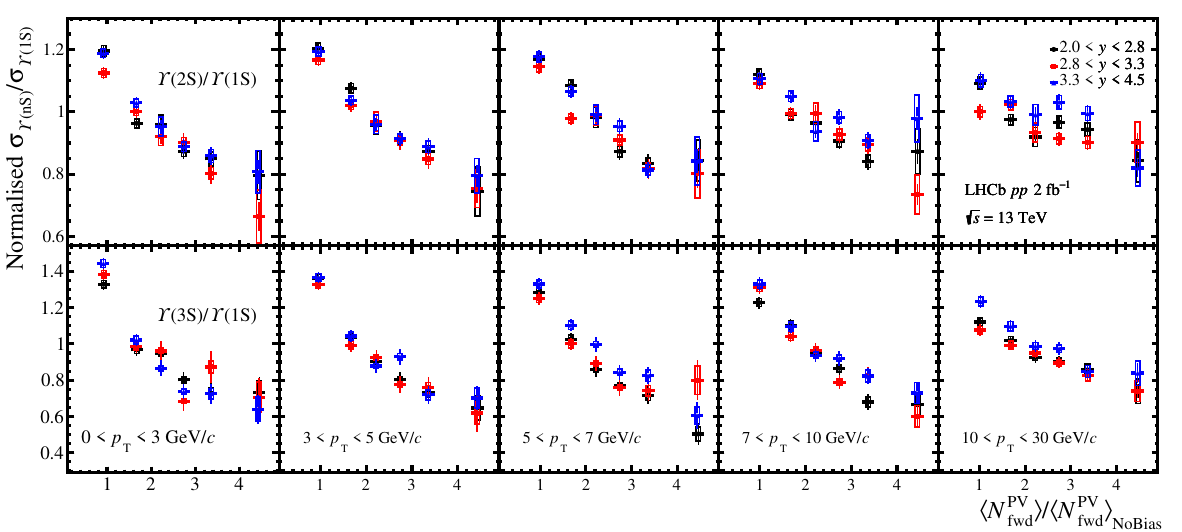}
\includegraphics[width=1.0\textwidth]{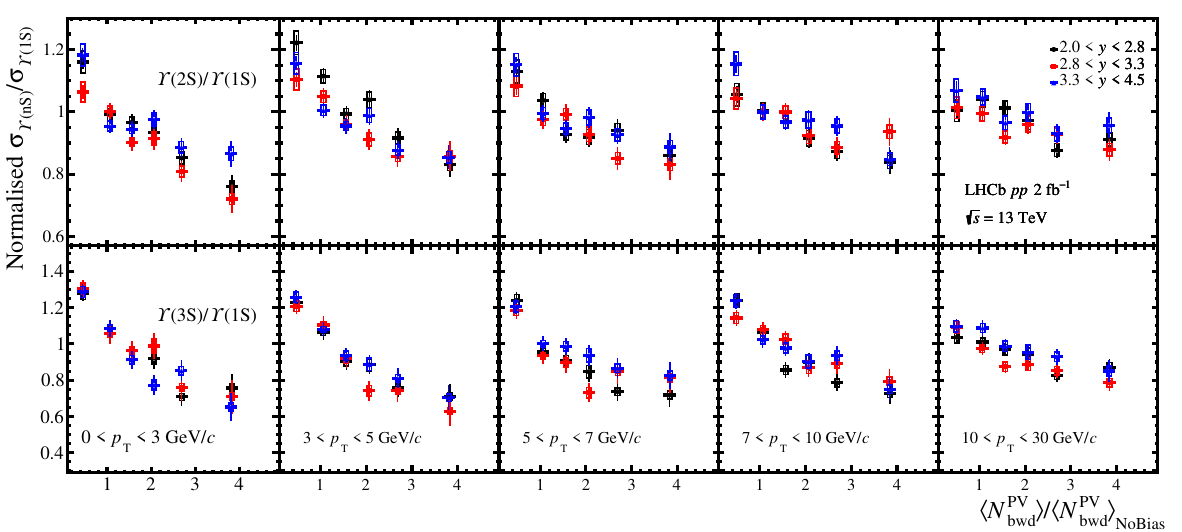}
\caption{Normalised double-differential production cross-section ratios as a function of self-normalised (top)~$N_{\mathrm{trk}}^{\mathrm{PV}}$, (middle)~$N_{\mathrm{fwd}}^{\mathrm{PV}}$ and (bottom)~$N_{\mathrm{bwd}}^{\mathrm{PV}}$ in different kinematic bins. The upper and lower row of each plot represents the $\mit\Upsilon(\mathrm{2}S)/\mit\Upsilon(\mathrm{1}S)$ and $\mit\Upsilon(\mathrm{3}S)/\mit\Upsilon(\mathrm{1}S)$ 
results, respectively.} 
\label{fig:ratiopvt}
\end{figure}

A more detailed multiplicity dependence of double-differential production ratios is shown in Fig.~\ref{fig:ratiopvt}. 
The overall trends are consistent with the multiplicity dependence observed in the integrated kinematic region $\pvt$ for \mbox{$2.0<y<4.5$} and \mbox{$0<\pt<30$ \gevc} shown in Fig.~\ref{fig:ratiot} and Fig.~\ref{fig:ratiofb}, respectively.  
Furthermore, within each rapidity bin, the multiplicity dependence of $\Yn{n}/\Yn{1}$ ratios in different \pt regions is similar and consistent with the rapidity-integrated results in different \pt bins as shown in Fig.~\ref{fig:ratioy}. 
This indicates that in the pseudorapidity interval $2.0<y<4.5$, the multiplicity dependence of $\Yn{n}/\Yn{1}$ ratios is roughly the same, which is consistent with the $\psi(2S)/\jpsi$ results~\cite{LHCb:2023xie}.

\clearpage
\subsection{Comparisons with other measurements}
\begin{figure}[!tb]
\centering
\begin{minipage}[t]{0.49\textwidth}
\centering
\includegraphics[width=1.0\textwidth]{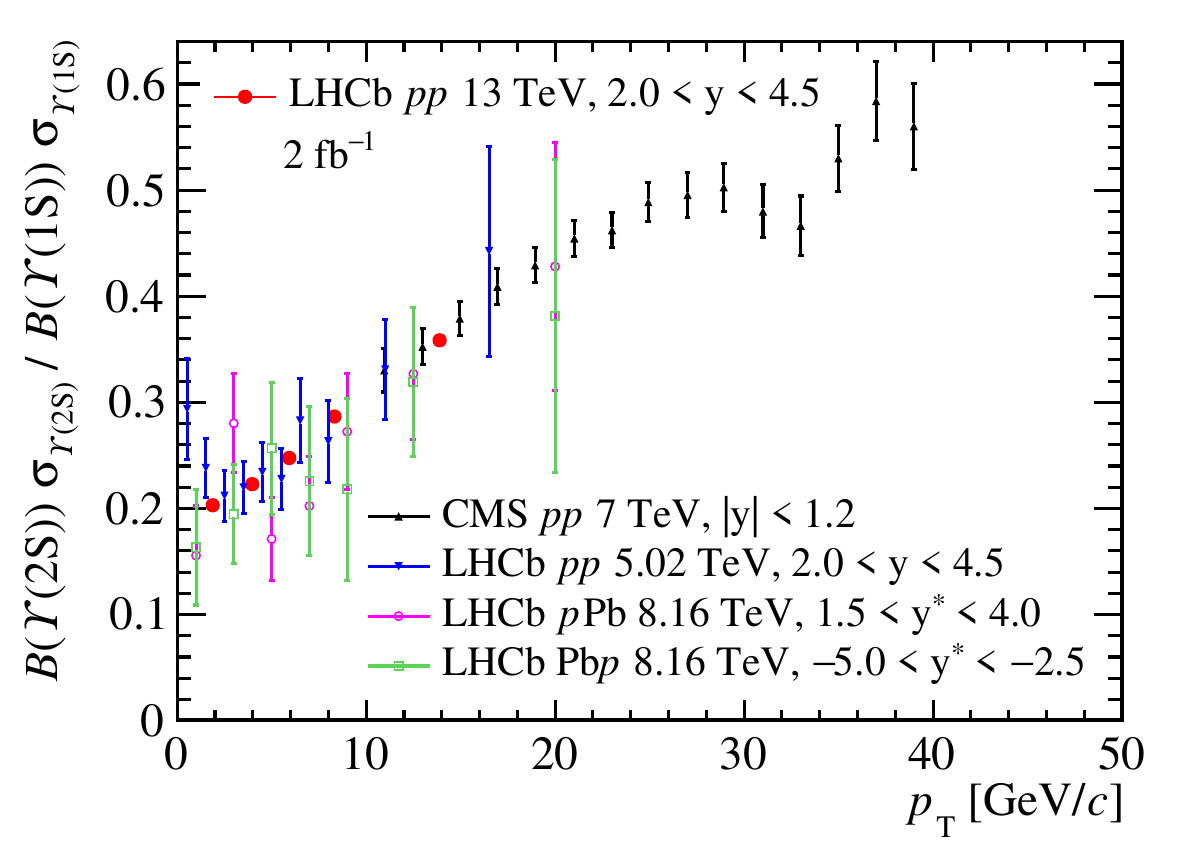}
\end{minipage}
\begin{minipage}[t]{0.49\textwidth}
\centering
\includegraphics[width=1.0\textwidth]{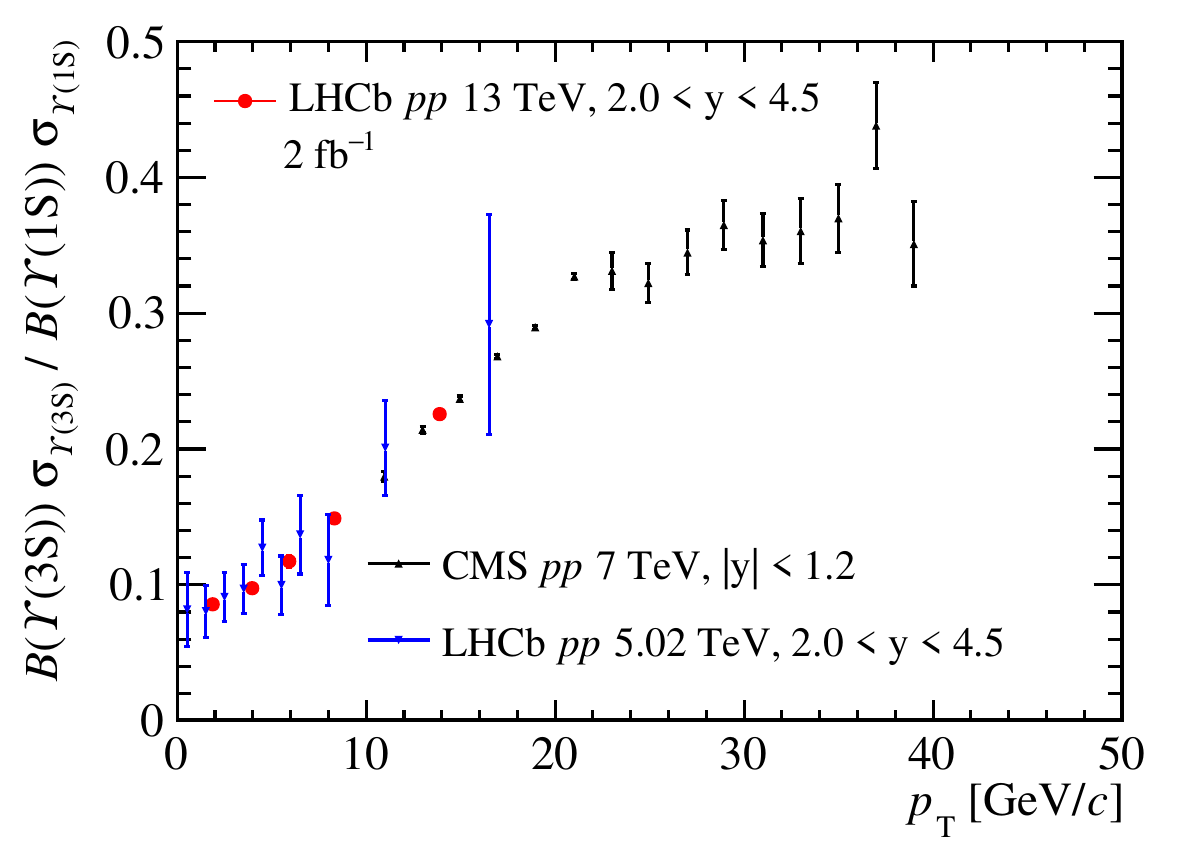}
\end{minipage}
\caption{The production times dimuon branching fraction ratio for (left)~$\mit\Upsilon(\mathrm{2}S)/\mit\Upsilon(\mathrm{1}S)$ and (right)~$\mit\Upsilon(\mathrm{3}S)/\mit\Upsilon(\mathrm{1}S)$ as a function of $\pt$ in different collision systems and experiments\cite{LHCb:2018psc,LHCb:2022byt,CMS:2015xqv}. Here, $y^*$ is the rapidity in the centre-of-mass frame. The error bars represent the quadratic sum of statistical and systematic uncertainties.} 
\label{fig:compt}
\end{figure}

Measurements of production ratios multiplied by $\BR(\Yn{n})/\BR(\Yn{1})$, where $n=2,3$, as a function of \pt, have been carried out in different collision systems and experiments\cite{LHCb:2018psc,LHCb:2022byt,CMS:2015xqv}. A comprehensive comparison is shown in Fig.~\ref{fig:compt}. The results presented in this paper show good agreement with the LHCb results measured in $pp$ collisions at 5\,\tev. They are also consistent with other measurements, but are more precise and significantly extend the rapidity region.

\section{Conclusion}
The production ratios of $\Yn{3}$ and $\Yn{2}$ to $\Yn{1}$ in $pp$ collisions at $\sqrt{s}=13\tev$ are studied using data collected by the LHCb experiment during 2018 and corresponding to an integrated luminosity of $2$\invfb.
 The normalised ratios are determined as functions of different multiplicity variables in bins of (\pt, $y$) and an integrated region over $2.0<y<4.5$ and $0<\pt<30 \gevc$. 
 A decreasing trend is observed for the normalised $\Yn{3}/\Yn{1}$ and $\Yn{2}/\Yn{1}$ ratios. 
 The $\Yn{3}$ state is found to be more suppressed, in line with other observations in larger collision systems. This hierarchy is found to be qualitatively consistent with the comover
model predictions, indicating that final-state interactions play an important role in bottomonia production in high-multiplicity events. 
The decreasing trend versus multiplicity is more significant when the multiplicity variable is measured in the rapidity region of the $\Yn{n}$ candidates. 
The decreasing trend in the low-\pt regions is more pronounced than in high-\pt regions. On the other hand, no significant dependence on rapidity is observed. Finally, the ratios as a function of \pt are compared with LHCb $pp$ and $p$Pb, and CMS results, and are in agreement within uncertainties.

%% file: acknowledgements.tex
\section*{Acknowledgements}
%
%
\noindent We express our gratitude to our colleagues in the CERN
accelerator departments for the excellent performance of the LHC. We
thank the technical and administrative staff at the LHCb
institutes.
We acknowledge support from CERN and from the national agencies:
CAPES, CNPq, FAPERJ and FINEP (Brazil); 
MOST and NSFC (China); 
CNRS/IN2P3 (France); 
BMBF, DFG and MPG (Germany); 
INFN (Italy); 
NWO (Netherlands); 
MNiSW and NCN (Poland); 
MCID/IFA (Romania); 
MICIU and AEI (Spain);
SNSF and SER (Switzerland); 
NASU (Ukraine); 
STFC (United Kingdom); 
DOE NP and NSF (USA).
We acknowledge the computing resources that are provided by CERN, IN2P3
(France), KIT and DESY (Germany), INFN (Italy), SURF (Netherlands),
PIC (Spain), GridPP (United Kingdom), 
CSCS (Switzerland), IFIN-HH (Romania), CBPF (Brazil),
and Polish WLCG (Poland).
We are indebted to the communities behind the multiple open-source
software packages on which we depend.
Individual groups or members have received support from
ARC and ARDC (Australia);
Key Research Program of Frontier Sciences of CAS, CAS PIFI, CAS CCEPP, 
Fundamental Research Funds for the Central Universities, 
and Sci. \& Tech. Program of Guangzhou (China);
Minciencias (Colombia);
EPLANET, Marie Sk\l{}odowska-Curie Actions, ERC and NextGenerationEU (European Union);
A*MIDEX, ANR, IPhU and Labex P2IO, and R\'{e}gion Auvergne-Rh\^{o}ne-Alpes (France);
AvH Foundation (Germany);
ICSC (Italy); 
Severo Ochoa and Mar\'ia de Maeztu Units of Excellence, GVA, XuntaGal, GENCAT, InTalent-Inditex and Prog. ~Atracci\'on Talento CM (Spain);
SRC (Sweden);
the Leverhulme Trust, the Royal Society
 and UKRI (United Kingdom).

%% file: Authorship_LHCb-PAPER-2024-038.tex
\centerline
{\large\bf LHCb collaboration}
\begin
{flushleft}
\small
R.~Aaij$^{38}$\lhcborcid{0000-0003-0533-1952},
A.S.W.~Abdelmotteleb$^{57}$\lhcborcid{0000-0001-7905-0542},
C.~Abellan~Beteta$^{51}$,
F.~Abudin{\'e}n$^{57}$\lhcborcid{0000-0002-6737-3528},
T.~Ackernley$^{61}$\lhcborcid{0000-0002-5951-3498},
A. A. ~Adefisoye$^{69}$\lhcborcid{0000-0003-2448-1550},
B.~Adeva$^{47}$\lhcborcid{0000-0001-9756-3712},
M.~Adinolfi$^{55}$\lhcborcid{0000-0002-1326-1264},
P.~Adlarson$^{82}$\lhcborcid{0000-0001-6280-3851},
C.~Agapopoulou$^{14}$\lhcborcid{0000-0002-2368-0147}
% ...

{\footnotesize \it

% ...
$^{14}$Universit{\'e} Paris-Saclay, CNRS/IN2P3, IJCLab, Orsay, France\\
% ...
$^{38}$Nikhef National Institute for Subatomic Physics, Amsterdam, Netherlands\\
% ...
$^{47}$Instituto Galego de F{\'\i}sica de Altas Enerx{\'\i}as (IGFAE), Universidade de Santiago de Compostela, Santiago de Compostela, Spain\\
% ...
$^{51}$Physik-Institut, Universit{\"a}t Z{\"u}rich, Z{\"u}rich, Switzerland\\
% ...
$^{55}$H.H. Wills Physics Laboratory, University of Bristol, Bristol, United Kingdom\\
% ...
$^{57}$Department of Physics, University of Warwick, Coventry, United Kingdom\\
% ...
$^{60}$School of Physics and Astronomy, University of Glasgow, Glasgow, United Kingdom\\
$^{61}$Oliver Lodge Laboratory, University of Liverpool, Liverpool, United Kingdom\\
% ...
$^{69}$Syracuse University, Syracuse, NY, United States\\
% ...
$^{82}$Department of Physics and Astronomy, Uppsala University, Uppsala, Sweden, associated to $^{60}$\\
% ...
}
\end{flushleft}